\documentclass[11pt,twoside]{article}
\usepackage{asp2006}
\usepackage{graphics}
\usepackage{graphicx}
\begin{document}
\title{On the Origin of Strong-Field Polarity Inversion Lines}
\author{B.~T. Welsch\altaffilmark{1} and Y. Li\altaffilmark{1}}

\altaffiltext{1}{Space Sciences Laboratory, University of California, 
7 Gauss Way, Berkeley, CA 94720-7450}

\begin{abstract}
Several studies have correlated observations of impulsive solar activity
--- flares and coronal mass ejections (CMEs) --- with the amount of
magnetic flux near strong-field polarity inversion lines (PILs) in
active regions' photospheric magnetic fields, as measured in
line-of-sight (LOS) magnetograms.  Practically, this empirical
correlation holds promise as a space weather forecasting tool.
Scientifically, however, the mechanisms that generate strong gradients
in photospheric magnetic fields remain unknown.  Hypotheses include:
the (1) emergence of highly twisted or kinked flux ropes, which
possess strong, opposite-polarity fields in close proximity; (2)
emergence of new flux in close proximity to old flux; and (3) flux
cancellation driven by photospheric flows acting fields that have
already emerged.  If such concentrations of flux near strong gradients
are formed by emergence, then increases in unsigned flux near strong
gradients should be correlated with increases in total unsigned
magnetic flux --- a signature of emergence.  Here, we analyze time
series of MDI line-of-sight (LOS) magnetograms from several dozen
active regions, and conclude that increases in unsigned flux near
strong gradients tend to occur during emergence, though strong
gradients can arise without flux emergence.  We acknowledge 
support from NSF-ATM 04-51438.
\end{abstract}

\section{Strong gradients across PILs}
\label{sec:intro}

It has been known for decades that flares and filament eruptions (which
form CMEs) originate along polarity inversion lines (PILs) of the radial
photospheric magnetic field.  In studies using photospheric vector
magnetograms, Falconer {\em et al.}~(2003, 2006)
\nocite{BW_Falconer2003,BW_Falconer2006} reported a strong correlation
between active region CME productivity and the total length of PILs with
strong potential transverse fields ($>150$ G) and strong gradients in
the LOS field (greater than 50 G Mm$^{-1}$).  They used a $\pm$2-day
temporal window for correlating magnetogram properties with CMEs. 
Falconer {\em et al.}~(2003) \nocite{BW_Falconer2003} noted that these
correlations remained essentially unchanged for ``strong gradient''
thresholds from 25 to 100 G Mm$^{-1}$.  Using more than 2500 MDI (LOS)
magnetograms, Schrijver (2007) \nocite{BW_Schrijver2007} found a strong
correlation between major (X- and M-class) flares and the total unsigned
magnetic flux near (within $\sim 15$ Mm) strong-field PILs --- defined,
in his work, as regions where oppositely signed LOS fields that exceed
150 G lie closer to each other than the instrument's $\sim$ 2.9 Mm
resolution.  Schrijver's (2007) effective gradient threshold, {$\sim$
100 G Mm$^{-1}$}, is stronger than that used by Falconer {\em et
al.}~(2003, 2006). \nocite{BW_Falconer2003,BW_Falconer2006}

Although these studies were published recently, the association
between flares and $\delta$ sunspots, which posses opposite-sign
umbrae within the same penumbra --- and therefore also possess
strong-field PILs --- has been well known for some time
\cite{BW_Kunzel1960,BW_Sammis2000}.  In particular, $\beta \gamma \delta$
spot groups are most likely to flare \cite{BW_Sammis2000}.  A $\beta
\gamma$ designation means no obvious north-south PIL is present in an
active region \cite{BW_Zirin1988}.

We note that Cui {\em et al.}~(2006) \nocite{BW_Cui2006} found that
the occurrence of flares is correlated with the maximum magnitude 
of the horizontal gradient in active region LOS magnetograms 
--- not just near PILs --- and that the correlation increases 
strongly for gradients stronger than $\sim$ 400 G Mm$^{-1}$.

One would expect the measures of CME- and flare- productivity
developed by both Falconer {\em et al.}~(2003,2006)
\nocite{BW_Falconer2003,BW_Falconer2006} Schrijver (2007)
\nocite{BW_Schrijver2007} to be larger for larger active regions.
Importantly, however, both studies showed that their measures of
flux near strong-field PILs is a better predictor of flare
productivity than total unsigned magnetic flux.  Evidently, more flux is
not, by itself, as significant a predictor of flares as more flux near
strong-field PILs.

These intriguing results naturally raise the question, 
``How do strong-field PILs form?''  For brevity, we
hereafter refer to strong-field PILs as SPILs. 

Schrijver (2007) \nocite{BW_Schrijver2007} contends that large SPILs form
primarily, if not solely, by emergence.  But he also noted that flux
emergence, by itself, does not necessarily lead to the formation of
SPILs.  Rather, a particular type of magnetic structure must emerge,
one containing a long SPIL at its emergence.  He suggests such
structures are horizontally oriented, filamentary currents.

Beyond the ``intact emergence'' scenario presented by Schrijver
(2007), \nocite{BW_Schrijver2007} other mechanisms can generate SPILs.
When new flux emerges in close proximity to old flux --- a common
occurrence \cite{BW_Harvey1993} --- SPILs can form along the boundaries
between old and new flux systems.  Converging motions in flux that has
already emerged can also generate SPILs.  If the convergence leads to
flux cancellation by some mechanism --- emergence of U loops,
submergence of inverse-U loops, or reconnective cancellation
\cite{BW_Welsch2006} --- then the total unsigned flux in the neighborhood
of the SPIL might decrease as the SPIL forms.  We note that, while
cancellation in already-emerged fields can occur via flux emergence
(from upward moving U-loops), the emergence of a new flux system
across the photosphere must increase the total unsigned flux that
threads the photosphere.  

If the emergence of new flux were primarily responsible for
SPILs, then a straightforward prediction would be that an increase in
total unsigned flux should be correlated with an increase in the
amount of unsigned flux near SPILs.  Hence, observations showing that
increases in the unsigned flux near SPILs frequently occur without a
corresponding increase in total unsigned flux would rule out new flux
emergence as the sole cause of these strong field gradients.

Our goal is to investigate the relationship between increases in the
amount of unsigned flux near SPILs with changes in unsigned flux in
the active regions containing the SPILs, to determine, if possible,
which processes generate SPILs.  

\section{Data}
\label{data}

From days-long time series of deprojected, 96-minute, full-disk MDI
magnetograms for $N_{AR} = $ 64 active regions, we computed the rates
of change of unsigned flux near SPILs, following the method described
by Schrijver (2007). \nocite{BW_Schrijver2007} We also computed the rates
of change of total unsigned line-of-sight magnetic flux these active
regions.

Our active region sample was chosen for use in a separate study of the
relationships between surface flows derived from magnetograms and
CMEs.  For the purposes of that study, we typically selected active
regions with a single, well defined PIL, for ease in identifying the
presence of shearing and/or converging flows some CME models employ
\cite{BW_Antiochos1999a,BW_Linker2001}.  The sample used here includes
regions from 1996 - 1998, and includes regions that did and did not
produce CMEs.  Some of our magnetograms image the same active region
as it rotated back onto the disk one or more times.  Also, some of our
selected regions are so decayed that they lack spots, and therefore
have no NOAA designation.

Here, we analyze $N_{\rm mag} = 4062$ magnetograms. Pixels more that
45$^\circ$ from disk center were ignored.  To convert the LOS field,
$B_{\rm LOS}$, to an estimated radial field, $B_R$, cosine corrections
were used, $B_R = B_{\rm LOS}/\cos(\Theta)$, where $\Theta$ is the
angle from disk center.

\begin{figure}[!ht]
\includegraphics[width=5.5in]{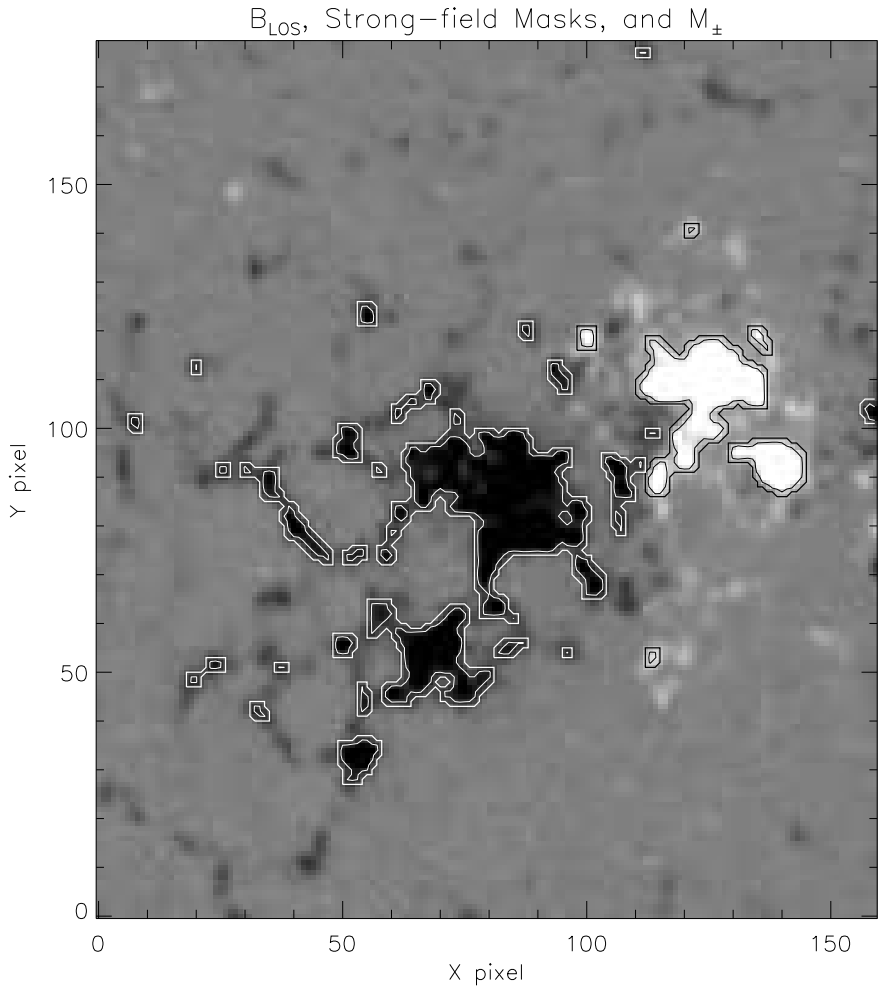}
  \caption{The background grayscale is typical, reprojected
  magnetogram; white is positive flux, black is negative flux.  The
  inner black and white contours enclose signed, strong-field masks ---
  regions where $B_R > 150$ G and $B_R < -150$ G (respectively). The
  outer black and white contours show the outlines of $M_\pm$,
  dilated bitmaps of the strong-field masks with a 3 $\times$ 3 kernel
  function.  For this magnetogram, the dilated bitmaps overlap at a
  single pixel, at $(x,y) =$ (112,91).}
\label{fig:uno}
\end{figure}

Triangulation was used to interpolate the $B_R$ data --- regularly
gridded in the plane-of-sky, but irregularly gridded in spherical
coordinates $(\theta,\phi)$ on the solar surface --- onto points
$(\theta',\phi')$ corresponding to a regularly gridded, Mercator
projection of the spherical surface.  This projection was adopted
because it is conformal (locally shape-preserving), necessary to
ensure displacements measured in the tracking study mentioned above
were not biased in direction.  For computing gradients, a conformal
projection is also appropriate.  The background grayscale in Figure
\ref{fig:uno} is a typical reprojected magnetogram.  We note that the
price of preserving shapes in the deprojection is distortion of scales;
but this can be easily corrected.

Each active region was tracked over 3 - 5 days, and cropped with a
moving window.  A list of tracked active regions and mpeg movies of
the active regions, are online, at 
\texttt{http://sprg.ssl.berkeley.edu/$\sim$yanli/lct/}~.

\section{Analysis methods}
\label{sec:methods}

To identify SPILs, we used the gradient identification technique of
Schrijver (2007).  \nocite{BW_Schrijver2007} For a magnetogram at time $t_i$,
binary positive/negative strong-field masks --- where
$B_R > 150$ G and $B_R < -150$ G, respectively --- were constructed,
then dilated by a (3x3) kernel to create dilated positive and negative
bitmaps, $M_\pm$.  These procedures are illustrated in Figure \ref{fig:uno}.
Regions of overlap, where $M_{\rm OL} = M_+ M_- = 1$, were identified
as SPILs.  In Figure \ref{fig:uno}, $M_{\rm OL} \ne 0$ for a single
pixel, at $(x,y) =$ (112,91).

\begin{figure}[!ht]
\includegraphics[width=5.5in]{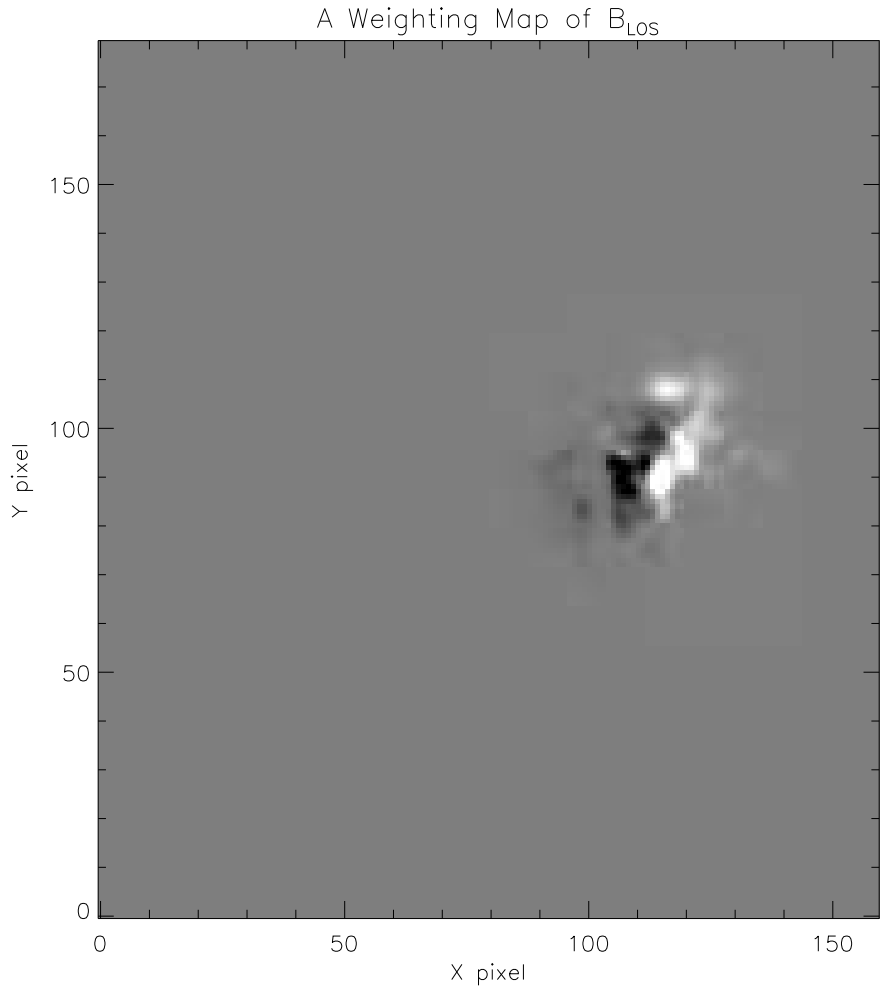}
  \caption{$B_R$ multiplied by a weighting map, $C_{MG}$. $R$ is the
  total unsigned field the window, $\sum |B_R| \, C_{MG}$. }
\label{fig:dos}
\end{figure}

To quantitatively define neighborhoods around SPILs, $M_{\rm OL}$ is
convolved with a normalized Gaussian,
\begin{equation}
G(u,v) = G_0^{-1} \exp(-[u^2+v^2]/2\sigma^2) 
\end{equation}
where $G_0 = \int du \int dv \exp(-[u^2+v^2]/2\sigma^2)$, and
$\sigma = 9$ pixels (corresponding to a FWHM $\simeq$ 15 Mm at disk center), to create 
``weighting maps,'' $C_{MG}$, where
\begin{equation}
C_{MG}(x,y) = {\rm convol}(M_{\rm OL}(x,y), G)
~. \end{equation}
 Figure \ref{fig:dos} shows the product of $B_R$ with such a weighting map.
Following Schrijver (2007), \nocite{BW_Schrijver2007} we totaled the
unsigned magnetic field over the weighting map to determine $R$, a
measure of the unsigned flux near SPILs,
\begin{equation}
R = \sum |B_R| \, C_{MG}  ~. \label{eqn:rdef}
\end{equation} 
From a sample of more than 2500 MDI magnetograms, Schrijver (2007) 
\nocite{BW_Schrijver2007} showed that $R$ is correlated with major (X- and
M-class) flares.

For each of the $N_R$ =1621 magnetograms with $R \ne 0$, 
we summed the weighted absolute magnetic field in the previous magnetogram, 
$B_R(t_{i-1})$, using the weighting map from $t_i$, to compute 
the backwards-difference $\Delta R$, 
\begin{equation}
\Delta R = \sum (|B_R(t_i)| - |B_R(t_{i-1})|) \, C_{MG}  
~.  \end{equation} 

We also computed the change in summed, unsigned field, 
\begin{equation}
\Delta {\cal B} = \sum (|B_R(t_i)| - |B_R(t_{i-1})|) ~, \end{equation}
to determine if new flux is emerging or if flux is canceling.  If new
flux is emerging, we expect $\Delta {\cal B} > 0$.  If flux is
canceling, we expect $\Delta {\cal B} < 0$.  Like Schrijver (2007),
\nocite{BW_Schrijver2007} we have opted to keep $R$ in units of flux
density; for simplicity, we also keep $\Delta {\cal B}$ in these same
units.

When the overlap map $M_{\rm OL}$ for $B_R(t_i)$ is identically zero, 
$R$ is also zero, and $\Delta R$ and $\Delta {\cal B}$ are not computed.

In \S \ref{sec:intro} we discussed processes that can cause changes in
$R.$ What processes can lead to $\Delta {\cal B} \ne 0$?  Emergence of
new flux or cancellation (both only happen at PILs) can make $\Delta
{\cal B} \ne 0$, and these processes are probably related to evolution
in $R$.  Flux can also cross into or out of the cropping window. Since
our cropping windows were selected to include essentially all of each
tracked active region's flux, systematic errors arising in this way are
expected to be small.  A more severe effect is the ``unipolar
appearance'' phenomenon characterized by Lamb {\em et al.} (2007),
\nocite{BW_Lamb2007} who found that the majority of newly detected flux
in the quiet sun is due to coalescence of previously existing, but
unresolved, single-polarity flux into concentrations large and strong
enough to detect.  While it is unclear if the conclusion reached by Lamb
{\em et al.} (2007) \nocite{BW_Lamb2007} for the quiet sun also applies
in active regions, this is plausible.  Moreover, much as flux can
``appear,'' flux can also disappear, via dispersive photospheric flows
or perhaps even molecular diffusivity.  Also, simultaneous emergence of
new flux and cancellation of existing flux can occur within the same
active region, masking the effects of both processes. Practically,
therefore, we can only refer to increases in unsigned flux as
``possible new flux emergence,'' and to decreases in unsigned flux as
``possible cancellation.''

\section{Results and conclusions}

\begin{figure}[!ht]
\includegraphics[width=5.5in]{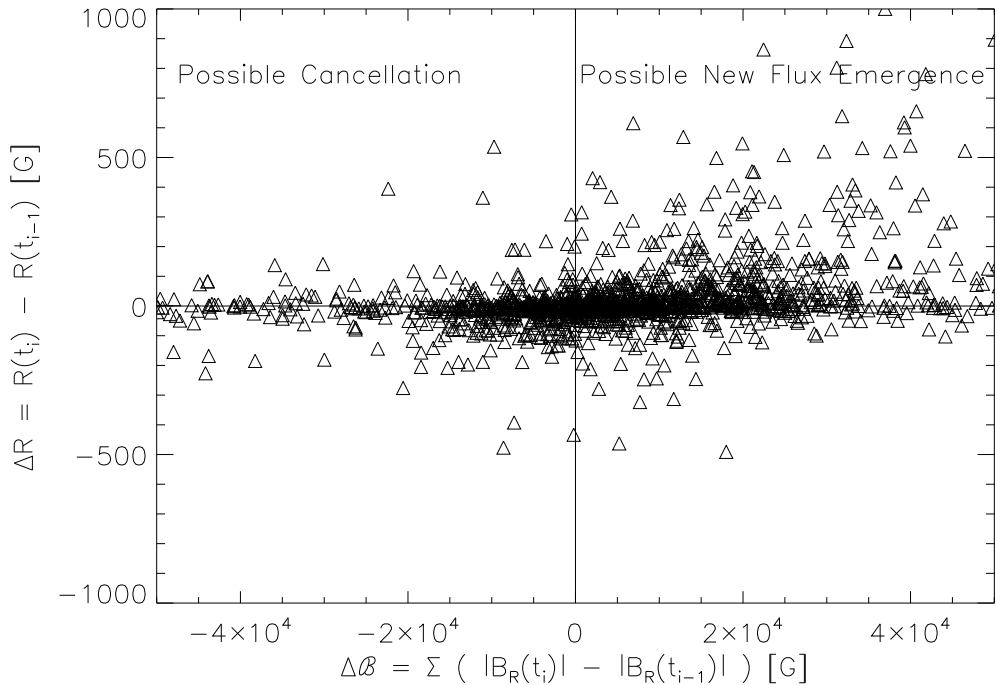}
  \caption{A scatter plot of changes in $R$ as a function of changes
    in ${\cal B}$.  Increases in $R$, the unsigned flux near SPILs,
    usually occur simultaneously with increases in the unsigned flux
    over the entire active region. Increases in $R$ only occur rarely
    when flux is decreasing, i.e., during cancellation.  For a
    breakdown of the data points in each quadrant, see Table
    \ref{tab:uno}.}
\label{fig:tres}
\end{figure}

In Figure \ref{fig:tres}, we show a scatter plot of changes in $R$ as
a function of changes in ${\cal B}$.  The plot does not show the full
range in $\Delta {\cal B}$, but the $\Delta R$ for outliers on the
horizontal axes are near zero.  One striking feature of the plot is
its flatness, i.e., that most changes in ${\cal B}$ are not associated
with any change in $R$.  In Table \ref{tab:uno}, we tabulated the data
points in each quadrant of this plot.  Clearly, increases in $R$, the
unsigned flux near SPILs, usually occur simultaneously with increases
in the unsigned flux over the entire active region. Increases in $R$
only occur less frequently when flux is decreasing, i.e., during 
cancellation.

\begin{table}
\caption{Breakdown of Flux Changes \label{tab:uno}}
\begin{tabular}{l|c|c}  
& $\Delta {\cal B} < 0$ & $\Delta {\cal B} > 0$ \\ \hline
$\Delta R > 0$ & 215 & 671 \\ \hline
$\Delta R < 0$ & 363 & 371 \\ 
\end{tabular}
\end{table}

We set out to answer the question, ``How do strong-field PILs form?''
We related changes in total, unsigned flux over whole active 
regions with changes in total, unsigned flux in subwindows of 
the same active regions --- defined by weighting maps.  
One might expect, therefore, that these quantities should be 
correlated, casting doubt about our ability to discrimintate 
between changes in total flux in active regions and in subwindows.  
If the two were strongly correlated, the excess of events with $\Delta
R > 0$ and $\Delta {\cal B} > 0$ might not be very meaningful.  In
fact, however, $\Delta R$ and $\Delta {\cal B}$ are poorly correlated:
the two have a linear correlation coefficient $r = 0.29$, and a
rank-order coefficient of $0.36.$ This suggests that the relationship
between increases in $R$ and increases in total, unsigned active
region flux is not an artifact of our approach.

Nonetheless, our active region sample is not ideally suited to address
the origin of SPILs, generally.  Our sample was not unbiased with
respect to active region morphology; we selected regions with
well-defined PILs.  In addition, our sample included some decayed
active regions that NOAA AR designations.  Consequently, we believe
that a follow-up study, with a much larger, unbiased sample of active
regions, is warranted.

With caveats, therefore, our study supports Schrijver's (2007)
\nocite{BW_Schrijver2007} contention that the emergence of new flux
creates the strong-field polarity inversion lines that he found to 
be correlated with flares.

\acknowledgements We acknowledge the support of NSF Grant NSF-ATM
04-51438.

\end{document}